\documentstyle[12pt,epsf]{article}
\begin{document}
\begin{center}
{\large\bf Covariant L-S Scheme for the effective $N^*NM$ couplings}

\vspace{1cm}

B. S. Zou$^{a,b,c,}$\footnote{zoubs@mail.ihep.ac.cn}, 
F. Hussain$^{c,}$\footnote{hussainf@ictp.trieste.it}

a) CCAST (World Laboratory), P.O.~Box 8730, Beijing 100080 \\
b) Institute of High Energy Physics, CAS,  P.~O.~Box 918(4), Beijing
100039, China\\
c) Abdus Salam International Centre for Theoretical Physics, Trieste, 
Italy\\

\date{\today}
\end{center}
\begin{abstract}
For excited nucleon states $N^*$ of arbitrary spin coupling to nucleon (N) and
meson (M), we propose a Lorentz covariant orbital-spin (L-S) scheme for the
effective $N^*NM$ couplings. To be used for the partial wave
analysis of various $N^*$ production and decay processes, it combines merits of
two conventional schemes, {\sl i.e.}, covariant effective Lagrangian approach
and multipole analysis with amplitudes expanded according to  
angular momentum L.
As examples, explicit formulae are given for $N^*\to N\pi$, $N^*\to N\omega$
and $\psi\to N^*\bar N$ processes which are under current experimental studies.
\end{abstract}


\section{Introduction}
The study of the nucleon and its excited states $N^*$ can provide us with critical
insights into the nature of QCD in the confinement domain \cite{Isgur1}. They are
the simplest system in which the three colors of QCD neutralize into colorless
objects and the essential nonabelian character of QCD is manifest. However our
present knowledge on the $N^*$ spectroscopy is still very poor, with information
coming almost entirely from the old generation of $\pi N$ experiments of more than
twenty years ago \cite{PDG} and with many fundamental
issues not well understood\cite{Capstick1}.
Considering its importance for the understanding of the nonperturbative QCD, much
effort has been devoted to the study of the $N^*$ spectrum. A series of new
experiments on $N^*$ physics with electromagnetic probes have been started at modern
facilities such as TJNAF\cite{Clas}, ELSA\cite{Elsa}, GRAAL\cite{Graal},
SPRING8\cite{Spring8} and BEPC\cite{BES1}.

Abundant data have been accumulated for various $N^*$ production
and decay channels at these facilities in last few years. Now an
important task facing us is to perform partial wave amplitude
analysis (PWA) of these data to extract properties of $N^*$
resonances, such as their spin-parity, mass, width, decay
branching ratios, and so on. For $\pi N$ or $\gamma N$ to
meson-nucleon final states, the most commonly used PWA formalism
is the multipole analysis with amplitudes expanded according to
angular momentum L of meson-nucleon
system\cite{Hohler,Ardnt,Tiator,Manley,Dytman}. This formalism is
usually written in the meson-nucleon CM system, not in a covariant
form, hence not very convenient to be used for multi-step chain
processes, such as $J/\psi\to N^*\bar N$ with $N^*$ further decaying to
meson-nucleon. For a multi-step chain process, the covariant
effective Lagrangian approach\cite{Nimai,Olsson,Lee,Liang,BES1} is
more convenient. In this approach, the effective $N^*NM$ couplings
are constructed by Rarita-Schwinger wave functions for particles
of arbitrary spin\cite{Rarita}, 4-momenta of involved particles,
Dirac $\gamma$ matrices, etc., with constraint of general
symmetries required by the strong interaction. A problem for this
approach is that the amplitude is usually a mixture of various
orbital angular momenta L. Hence the usual centrifugal barrier
(Blatt-Weisskopf) factor\cite{Manley, Chung}, commonly used in
multipole analysis and mesonic decays, cannot be used here since
the barrier factor is L-dependent. Instead vertex form factors
with exponential form or other forms are used in the effective
Lagrangian approach. This makes comparison to results from usual
multipole approach very difficult.

In this paper we propose a covariant L-S Scheme for the effective
$N^*NM$ couplings to be used for the partial wave analysis of
$N^*$ data. In this scheme, the amplitudes are expanded according
to the orbital angular momentum L of two decay products, meanwhile
Lorentz invariant. Hence it combines the merits of multipole
analysis and the effective Lagrangian approach.

\section{General Formalism}

In our construction of the covariant L-S Scheme for the effective $N^*NM$ couplings,
we need to combine some knowledge from the covariant tensor formalism for meson
decays\cite{Chung} and covariant wave functions for hadrons of arbitrary
spin\cite{Hussain}.

For a given hadronic decay process $A\to BC$, in the L-S scheme on
hadronic level, the initial state is described by its 4-momentum
$P_\mu$ and its spin state $\bf S_A$; the final state is described
by the relative orbital angular momentum state of BC system $\bf
L_{BC}$ and their spin states ($\bf S_B$, $\bf S_C$).

The spin states ($\bf S_A$, $\bf S_B$, $\bf S_C$) can be well
represented by the relativistic Rarita-Schwinger spin wave
functions for particles of arbitrary
spin\cite{Rarita,Chung,Zhujj,Liang}.  The spin-$1\over 2$
wavefunction is the standard Dirac spinor u(p,s) or v(p,s) and the
spin-1 wave function is the standard spin-1 polarization
four-vector $\varepsilon^\mu(p,s)$ for particle with momentum p
and spin projection $s$.
\begin{equation}
\sum_{s=0,\pm 1}\varepsilon_\mu(p,s)\varepsilon^*_\nu(p,s)
=-g_{\mu\nu}+{p_\mu p_\nu\over p^2}\equiv\tilde g_{\mu\nu}(p).
\end{equation}
Spin wave functions for particles of higher spins are
constructed from these two basic spin wave functions with C-G coefficients
$(j_1,j_{1z};j_2,j_{2z}|j,j_z)$ as the following:
\begin{equation}
\varepsilon_{\mu_1\mu_2\cdots\mu_n}(p,n,s)
=\sum_{s_{n-1},s_n}(n-1,s_{n-1};1,s_n|n,s)
\varepsilon_{\mu_1\mu_2\cdots\mu_{n-1}}(p,n-1,s_{n-1})\varepsilon_{\mu_n}(p,s_n)
\end{equation}
for a particle with integer spin $n\ge 2$, and
\begin{equation}
u_{\mu_1\mu_2\cdots\mu_n}(p,n+{1\over 2},s)
=\sum_{s_n,s_{n+1}}(n,s_n;{1\over 2},s_{n+1}|n+{1\over 2},s)
\varepsilon_{\mu_1\mu_2\cdots\mu_n}(p,n-1,s_n)u(p,s_{n+1})
\end{equation}
for a particle with half integer spin $n+{1\over 2}$ of $n\ge 1$.

The orbital angular momentum $\bf L_{BC}$ state can be represented by covariant
tensor wave functions $\tilde t^{(L)}_{\mu_1\cdots\mu_L}$ as the same as for meson
decay\cite{Chung}. Define $r=p_B-p_C$, then
\begin{eqnarray}
\tilde t^{(0)} &=& 1 , \\
\tilde t^{(1)}_\mu &=& \tilde g_{\mu\nu}(p_A)r^\nu\equiv\tilde r_\mu ,\\
\tilde t^{(2)}_{\mu\nu} &=& \tilde r_\mu\tilde r_\nu
-{1\over 3}(\tilde r\cdot\tilde r)\tilde g_{\mu\nu}, \\
\tilde t^{(3)}_{\mu\nu\lambda} &=& \tilde r_\mu\tilde r_\nu\tilde r_\lambda
-{1\over 5}(\tilde r\cdot\tilde r)(\tilde g_{\mu\nu}\tilde r_\lambda
+\tilde g_{\nu\lambda}\tilde r_\mu+\tilde g_{\lambda\mu}\tilde r_\nu),\\
 & & \cdots  \nonumber
\end{eqnarray}

In the L-S scheme, we need to use the conservation relation of total angular
momentum:
\begin{equation}
\label{L}
{\bf S_A} = {\bf S_B}+{\bf S_C}+{\bf L_{BC}}\quad or \quad -{\bf S_A} +
{\bf S_B}+{\bf S_C}+{\bf L_{BC}}=0.
\end{equation}
Comparing with the pure meson case\cite{Chung}, here for $N^*NM$ couplings
we need to introduce the concept of relativistic total spin of two fermions.

For the case of A as a meson, B as $N^*$ with spin $n+{1\over 2}$ and C as $\bar N$
with spin-$1\over 2$, the total spin of BC ($\bf S_{BC}$) can be either $n$ or
$n+1$.  The two $\bf S_{BC}$ states can be represented as
\begin{eqnarray}
\psi^{(n)}_{\mu_1\cdots\mu_n} &=& \bar
u_{\mu_1\cdots\mu_n}(p_B,s_B)\gamma_5v(p_C,s_C),\\
\Psi^{(n+1)}_{\mu_1\cdots\mu_{n+1}} &=& \bar
u_{\mu_1\cdots\mu_n}(p_B,s_B)(\gamma_{\mu_{n+1}}-{r_{\mu_{n+1}}\over m_A+m_B+m_C})
v(p_C,s_C) 
\nonumber\\
& & +(\mu_1\leftrightarrow\mu_{n+1}) + \cdots + (\mu_n\leftrightarrow\mu_{n+1})
\end{eqnarray}
for $\bf S_{BC}$ of $n$ and $n+1$,
respectively. As a special case of $n=0$, we have
\begin{eqnarray}
\psi^{(0)} &=& \bar u(p_B,s_B)\gamma_5v(p_C,s_C) , \\
\Psi^{(1)}_\mu &=& \bar u(p_B,s_B)(\gamma_\mu-{r_\mu\over m_A+m_B+m_C}) 
v(p_C,s_C) .
\end{eqnarray}
Here $r_\mu$ term is necessary to cancel out the $\bf\hat p$-dependent component in 
the simple $\bar u\gamma_\mu v$ expression. In the A at-rest system, we have
\begin{eqnarray}
\psi^{(0)} &=& C_\psi(-1)^{{1\over 2}-s_C}\delta_{s_B(-s_C)},\\
\Psi^{(1)}_i &=& C_\Psi (-1)^{{1\over 2}-s_C}
\chi^\dagger_{s_B}{\bf\sigma}_i\chi_{-s_C}
\end{eqnarray}
with two-component Pauli spinors $\chi^\dagger_{1/2}=(1,0)$ and
$\chi^\dagger_{-1/2}=(0,1)$,  and
\begin{eqnarray}
C_\psi &=& {(E_B+m_B)(E_C+m_C)+{\bf p^2_C}\over
\sqrt{2m_B2m_C(E_B+m_B)(E_C+m_C)}} , \\
C_\Psi &=& \sqrt{(E_B+m_B)(E_C+m_C)\over 2m_B2m_C}
\left(1+{{\bf p^2_C}\over (E_B+m_B)(E_C+m_C)}\right) .
\end{eqnarray}
In the non-relativistic limit, both $C_\psi$ and $C_\Psi$ are equal to 1.
Generally both of them have some smooth dependence on the magnitude of momentum.
But both $\psi^{(0)}$ and $\Psi^{(1)}_\mu$ have no dependence on the direction of
the momentum $\bf\hat p$, hence correspond to pure spin states with the total spin
of 0 and 1, respectively.

For the case of A as $N^*$ with spin
$n+{1\over 2}$, B as $N$ and C as a meson, one needs to couple $\bf -S_A$ and
$\bf S_B$ first to get $\bf S_{AB}\equiv -S_A+S_B$ states, which are
\begin{eqnarray}
\phi^{(n)}_{\mu_1\cdots\mu_n} &=& \bar
u(p_B,s_B)u_{\mu_1\cdots\mu_n}(p_A,s_A),\\
\Phi^{(n+1)}_{\mu_1\cdots\mu_{n+1}} &=& \bar
u(p_B,s_B)\gamma_5\tilde\gamma_{\mu_{n+1}}u_{\mu_1\cdots\mu_n}(p_A,s_A) +
(\mu_1\leftrightarrow\mu_{n+1}) + \cdots + (\mu_n\leftrightarrow\mu_{n+1})
\nonumber\\
\end{eqnarray}
for $\bf S_{AB}$ of $n$ and $n+1$, respectively.
\begin{eqnarray}
\phi^{(0)} &=& \bar u(p_B,s_B)u(p_A,s_A) , \\
\Phi^{(1)}_\mu &=& \bar u(p_B,s_B)\gamma_5\tilde\gamma_\mu u(p_A,s_A) 
\end{eqnarray}
with $\tilde\gamma_\mu=\tilde g_{\mu\nu}(p_A)\gamma^\nu$.
In the A ($N^*$) at-rest system, we have
\begin{eqnarray}
\phi^{(0)} &=& \sqrt{(E_A+m_A)(E_B+m_B)\over 2m_A2m_B}\delta_{s_As_B},\\
\Phi^{(1)}_i &=& -\sqrt{(E_A+m_A)(E_B+m_B)\over 2m_A2m_B}
\chi^\dagger_{s_B}{\bf\sigma}_i\chi_{s_A}.
\end{eqnarray}
Both have no dependence on the direction of the momentum $\bf\hat p$.

In effective Lagrangian approaches, the effective $N^*NM$
couplings are constructed by $p_A$, $r$, $g^{\mu\nu}$,
$\gamma^\mu$, $u$ or $v$, and may be mixture of various orbital
angular momentum states. In our proposed covariant L-S scheme, the
effective $N^*NM$ couplings should be composed of $p_A$, $\tilde
t^{(L)}$, $g^{\mu\nu}$, $\epsilon_{\alpha\beta\gamma\delta}$ (the
full anti-symmetric tensor), $\psi$ ($\Psi$) or $\phi$ ($\Phi$),
corresponding to a pure orbital angular momentum L state. Then the
procedure for constructing the effective $N^*NM$ couplings is very
similar to the case for pure mesons\cite{Chung}. First the parity
should be conserved, which means
\begin{equation}
\label{parity}
\eta_A=\eta_B\eta_C(-1)^{L}
\end{equation}
where $\eta_A$, $\eta_B$ and $\eta_C$ are the intrinsic parities
of particles A, B and C, respectively. From this relation, one
knows whether L should be even or odd. Then from Eq.(\ref{L}) one
can figure out how many different L-S combinations, which
determine the number of independent couplings. For a final state
with orbital angular momentum of L, $\tilde t^{(L)}$ should appear
once in the effective coupling without any other $\tilde t$ or
$r$. This will guarantee a pure L final state. Then one can easily
put into the Blatt-Weisskopf centrifugal barrier factor for each
effective coupling with L final state if one wishes. We shall show
the concrete procedure by examples in the following section.

\section{Examples}

We shall start with the simplest case for $N^*\to N\pi$ process, then for $N^*\to
N\omega$ and $\psi\to N^*\bar N$ where $\psi$ can be $J/\psi$ or $\psi'$ or any
other heavy vector mesons.

\subsection{$N^*\to N\pi$}

For $N^*\to N\pi$, it is well known that only one possible L-S coupling for the
$N\pi$ final state of each $N^*$ decay. Since the nucleon has spin-parity
${1\over 2}^+$
and pion has spin-parity $0^-$, $N^*({1\over 2}^+)$ can only decay to $N\pi$ in
P-wave with $S_{AB}=1$ to make $-{\bf S_A} + {\bf S_B}+{\bf S_C}+{\bf L_{BC}}={\bf
S_{AB}}+{\bf L_{BC}}=0$ meanwhile satisfying parity conservation relation
Eq.(\ref{parity}). Similarly we have $N^*({1\over 2}^-)\to N\pi$ in S-wave with
$S_{AB}=0$;
$N^*({3\over 2}^+)\to N\pi$ in P-wave with $S_{AB}=1$; $N^*({3\over 2}^-)\to N\pi$
in D-wave; $N^*({5\over 2}^+)\to N\pi$ in F-wave with $S_{AB}=3$;
$N^*({5\over 2}^-)\to N\pi$ in D-wave with $S_{AB}=2$;
$N^*({7\over 2}^+)\to N\pi$ in F-wave with $S_{AB}=3$;
$N^*({7\over 2}^-)\to N\pi$ in G-wave with $S_{AB}=4$; and so on. Then the effective
$N^*N\pi$ couplings in the covariant L-S scheme are
\begin{eqnarray}
N^*({1\over 2}^+)\to N\pi: \quad & & \Phi^{(1)}_\mu\tilde t^{(1)\mu},  \\
N^*({1\over 2}^-)\to N\pi: \quad & & \phi^{(0)}\tilde t^{(0)},   \\
N^*({3\over 2}^+)\to N\pi: \quad & & \phi^{(1)}_\mu\tilde t^{(1)\mu}, \\
N^*({3\over 2}^-)\to N\pi: \quad & & \Phi^{(2)}_{\mu\nu}\tilde t^{(2)\mu\nu},  \\
N^*({5\over 2}^+)\to N\pi: \quad & & \Phi^{(3)}_{\mu\nu\lambda}\tilde
t^{(3)\mu\nu\lambda}, \\
N^*({5\over 2}^-)\to N\pi: \quad & & \phi^{(2)}_{\mu\nu}\tilde t^{(2)\mu\nu},  \\
N^*({7\over 2}^+)\to N\pi: \quad & & \phi^{(3)}_{\mu\nu\lambda}\tilde
t^{(3)\mu\nu\lambda}, \\
N^*({7\over 2}^-)\to N\pi: \quad & & \Phi^{(4)}_{\mu\nu\lambda\delta}\tilde
t^{(4)\mu\nu\lambda\delta}.
\end{eqnarray}
Here for simplicity we omit the vertex form factors.
With properties of Rarita-Schwinger wave functions
\begin{equation}
\gamma^{\mu_i}u_{\cdots\mu_i\cdots}=0 \quad and \quad
p^{\mu_i}u_{\cdots\mu_i\cdots}(p,s)=0
\end{equation}
one can easily get the relation between the covariant L-S couplings and
the usual effective Lagrangian ones
\begin{eqnarray}
N^*({1\over 2}^+)\to N\pi: \quad & & \Phi^{(1)}_\mu\tilde t^{(1)\mu}=
\bar u_N\gamma_5\gamma_\mu u_* p^\mu_\pi\cdot C_\Phi, \\
N^*({1\over 2}^-)\to N\pi: \quad & & \phi^{(0)}\tilde t^{(0)}=\bar u_Nu_*\cdot 1,
\\
N^*({3\over 2}^+)\to N\pi: \quad & & \phi^{(1)}_\mu\tilde t^{(1)\mu}
=\bar u_Nu_{*\mu}p^\mu_\pi\cdot 2, \\
N^*({3\over 2}^-)\to N\pi: \quad & & \Phi^{(2)}_{\mu\nu}\tilde t^{(2)\mu\nu}
=\bar u_N\gamma_5\gamma_\mu u_{*\nu} p^\mu_\pi p^\nu_\pi\cdot 4C_\Phi,  \\
N^*({5\over 2}^+)\to N\pi: \quad & & \Phi^{(3)}_{\mu\nu\lambda}\tilde
t^{(3)\mu\nu\lambda} = \bar u_N\gamma_5\gamma_\mu u_{*\nu\lambda} p^\mu_\pi
p^\nu_\pi p^\lambda_\pi\cdot 12C_\Phi, \\
N^*({5\over 2}^-)\to N\pi: \quad & & \phi^{(2)}_{\mu\nu}\tilde t^{(2)\mu\nu}
=\bar u_Nu_{*\mu\nu}p^\mu_\pi p^\nu_\pi\cdot 4,  \\
N^*({7\over 2}^+)\to N\pi: \quad & & \phi^{(3)}_{\mu\nu\lambda}\tilde
t^{(3)\mu\nu\lambda}=\bar u_Nu_{*\mu\nu\lambda}p^\mu_\pi p^\nu_\pi
p^\lambda_\pi\cdot 8, \\
N^*({7\over 2}^-)\to N\pi: \quad & & \Phi^{(4)}_{\mu\nu\lambda\delta}\tilde
t^{(4)\mu\nu\lambda\delta}=\bar u_N\gamma_5\gamma_\mu u_{*\nu\lambda\delta}
p^\mu_\pi p^\nu_\pi p^\lambda_\pi p^\delta_\pi\cdot 48C_\Phi
\end{eqnarray}
with
\begin{equation}
C_\Phi=(1+{m_N\over m_*}-{m^2_\pi\over m^2_*+m_*m_N}),
\end{equation}
$u_N$, $u_*$ the Rarita-Schwinger wave functions of N and $N^*$, respectively;
$m_N$, $m_*$ the mass of N and $N^*$, respectively; $p_\pi$ the four-momentum of the
pion. We see the two approaches are equivalent here up to some constants or a
smooth $m_*$ dependent factor $C_\Phi$. This is
because for any $N^*\to N\pi$ process there is only one possible L-S coupling and
hence only one independent coupling.

\subsection{$N^*\to N\omega$}

Unlike pion with spin 0, here $\omega$ has spin 1. For $N^*$ with spin $1\over 2$
there are two independent L-S couplings conserving parity (\ref{parity}) and total
angular momentum (\ref{L}); for $N^*$ with spin larger than $1\over 2$, there are
three independent L-S couplings. Here we list them for $N^*$ with spin up to $7\over
2$.
\begin{eqnarray}
    &(S_C, S_{AB}, L_{BC})&: \quad {\bf S_{AB}}+{\bf S_C}+{\bf L_{BC}}=0
\nonumber\\
N^*({1\over 2}^+)\to N\omega &(1,0,1)&: \quad \phi^{(0)}\varepsilon^*_\mu\tilde
t^{(1)\mu}, \\
&(1,1,1)&: \quad 
i\Phi^{(1)}_\mu\epsilon^{\mu\nu\lambda\sigma}\varepsilon^*_\nu\tilde
t^{(1)}_\lambda\hat p_{*\sigma}, \\
N^*({1\over 2}^-)\to N\omega &(1,1,0)&: \quad \Phi^{(1)}_\mu\varepsilon^{*\mu},   \\
&(1,1,2)&: \quad \Phi^{(1)}_\mu\varepsilon^*_\nu\tilde t^{(2)\mu\nu} ,\\
N^*({3\over 2}^+)\to N\omega &(1,1,1)&: \quad
i\phi^{(1)}_\mu\epsilon^{\mu\nu\lambda\sigma}\varepsilon^*_\nu\tilde
t^{(1)}_\lambda\hat p_{*\sigma}, \\
&(1,2,1)&: \quad \Phi^{(2)}_{\mu\nu}\varepsilon^{*\mu}\tilde t^{(1)\nu}, \\
&(1,2,3)&: \quad \Phi^{(2)}_{\mu\nu}\varepsilon^*_\lambda\tilde
t^{(3)\mu\nu\lambda},
\\
N^*({3\over 2}^-)\to N\omega &(1,1,0)&: \quad \phi^{(1)}_\mu\varepsilon^{*\mu}, \\
&(1,1,2)&: \quad \phi^{(1)}_\mu\varepsilon^*_\nu\tilde t^{(2)\mu\nu}, \\
&(1,2,2)&: \quad i\Phi^{(2)}_{\mu\alpha}\epsilon^{\mu\nu\lambda\sigma}
\varepsilon^*_\nu\tilde t^{(2)\alpha}_\lambda\hat p_{*\sigma}, \\
N^*({5\over 2}^+)\to N\omega &(1,2,1)&: \quad
\phi^{(2)}_{\mu\nu}\varepsilon^{*\mu}\tilde t^{(1)\nu}, \\
&(1,2,3)&: \quad \phi^{(2)}_{\mu\nu}\varepsilon^*_\lambda\tilde
t^{(3)\mu\nu\lambda}, \\
&(1,3,3)&: \quad 
i\Phi^{(3)}_{\mu\alpha\beta}\epsilon^{\mu\nu\lambda\sigma}
\varepsilon^*_\nu\tilde t^{(3)\alpha\beta}_\lambda\hat p_{*\sigma}, \\
N^*({5\over 2}^-)\to N\omega &(1,2,2)&: \quad
i\phi^{(2)}_{\mu\alpha}\epsilon^{\mu\nu\lambda\sigma}
\varepsilon^*_\nu\tilde t^{(2)\alpha}_\lambda\hat p_{*\sigma}, \\
&(1,3,2)&: \quad \Phi^{(3)}_{\mu\nu\lambda}\varepsilon^{*\mu}\tilde
t^{(2)\nu\lambda}, \\
&(1,3,4)&: \quad \Phi^{(3)}_{\mu\nu\lambda}\varepsilon^*_\sigma\tilde
t^{(4)\mu\nu\lambda\sigma}, \\
N^*({7\over 2}^+)\to N\omega &(1,3,3)&: \quad
i\phi^{(3)}_{\mu\alpha\beta}\epsilon^{\mu\nu\lambda\sigma}
\varepsilon^*_\nu\tilde t^{(3)\alpha\beta}_\lambda\hat p_{*\sigma}, \\
&(1,4,3)&: \quad \Phi^{(4)}_{\mu\nu\lambda\sigma}\varepsilon^{*\mu}\tilde
t^{(3)\nu\lambda\sigma}, \\
&(1,4,5)&: \quad \Phi^{(4)}_{\mu\nu\lambda\sigma}\varepsilon^*_\delta\tilde
t^{(5)\mu\nu\lambda\sigma\delta}, \\
N^*({7\over 2}^-)\to N\omega &(1,3,2)&: \quad
\phi^{(3)}_{\mu\nu\lambda}\varepsilon^{*\mu}\tilde t^{(2)\nu\lambda}, \\
&(1,3,4)&: \quad \phi^{(3)}_{\mu\nu\lambda}\varepsilon^*_\sigma\tilde
t^{(4)\mu\nu\lambda\sigma}, \\
&(1,4,4)&: \quad
i\Phi^{(4)}_{\mu\alpha\beta\gamma}\epsilon^{\mu\nu\lambda\sigma}
\varepsilon^*_\nu\tilde t^{(4)\alpha\beta\gamma}_\lambda\hat p_{*\sigma}.
\end{eqnarray}
where $\hat p_{*\sigma}=p_{*\sigma}/m_*$.  In the $N^*$ at-rest system,
$\hat p_*=(1,0,0,0)$; $\epsilon^{\mu\nu\lambda\sigma}S_\mu L_\nu J_\lambda
\hat p_{*\sigma}=({\bf S}\times {\bf L})\cdot {\bf J}$ is the standard 
form for forming a total angular momentum $|{\bf J}|=1$ from two other 
angular momenta (S,L) of absolute value 1. In the covariant L-S tensor 
formalism, for S-L-J coupling, if S+L+J is an odd number, then the
$\epsilon^{\mu\nu\lambda\sigma}\hat p_{A\sigma}$ is needed.   
These are the only possible independent couplings because the fact that
$p_{*\sigma}t^{(n)\sigma\mu...} = 0$, $p_{*\sigma}\phi^{(n)\sigma\mu...} = 
0$ and $p_{*\sigma}\Phi^{(n)\sigma\mu...} = 0$.
The corresponding couplings from the simple effective Lagrangian approach
are give in Ref.\cite{Liang}. They have the same number of independent couplings
as here and are linear combinations of couplings here.
For example, for $N^*{3\over 2}^-\to N\omega$, the full amplitude
in the covariant L-S scheme is
\begin{equation}
A=g_1\phi^{(1)}_\mu\varepsilon^\mu+g_2\phi^{(1)}_\mu\varepsilon_\nu\tilde
t^{(2)\mu\nu}+g_3i\Phi^{(2)}_{\mu\alpha}\epsilon^{\mu\nu\lambda\sigma}
\varepsilon_\nu\tilde t^{(2)\alpha}_\lambda p_{*\sigma}
\end{equation}
with vertex form factors $g_1$, $g_2$ and $g_3$, while in the simple effective
Lagrangian approach\cite{Liang} is
\begin{equation}
A=f_1\bar u_N u_{*\mu}\varepsilon^\mu + f_2 \bar u_N \gamma_\nu u_{*\mu}p^\mu_N 
\varepsilon^\nu +f_3 \bar u_N u_{*\mu} p^\mu_\omega \varepsilon_\nu p^\nu_N
\end{equation}
with vertex form factors $f_1$, $f_2$ and $f_3$. These Vertex form factors are
smooth functions of $m_*$ with practically constant $m_N$ and $m_\omega$; they have
no dependence on angular variable.  With some simple algebra and the 
following
identity\cite{Korner}:
\begin{eqnarray}
i\epsilon_{\mu abc} &=&
\gamma_5(\gamma_\mu\gamma_a\gamma_b\gamma_c-\gamma_\mu\gamma_a g_{bc}
+\gamma_\mu\gamma_b g_{ac}-\gamma_\mu\gamma_c g_{ab} \nonumber\\
& & -\gamma_a\gamma_b g_{\mu c} + \gamma_a\gamma_c g_{\mu b} -\gamma_b\gamma_c
g_{\mu a} + g_{\mu a}g_{bc} - g_{\mu b}g_{ac} + g_{\mu c}g_{ab}) ,
\end{eqnarray}
we have \begin{eqnarray}
\phi^{(1)}_\mu\varepsilon^\mu &=& \bar u_N u_{*\mu}\varepsilon^\mu ,  \\
\phi^{(1)}_\mu\varepsilon_\nu\tilde t^{(2)\mu\nu} &=& 
2(-1+{m^2_N-m^2_\omega\over m^2_*})\bar u_N u_{*\mu} p^\mu_\omega \varepsilon_\nu
p^\nu_N +{1\over 3}{\bf r}^2 \bar u_N u_{*\mu}\varepsilon^\mu ,\\
i\Phi^{(2)}_{\mu\alpha}\epsilon^{\mu\nu\lambda\sigma}
\varepsilon_\nu\tilde t^{(2)\alpha}_\lambda\hat p_{*\sigma} &=&
2(-3+{m^2_N-m^2_\omega\over m^2_*})\bar u_N u_{*\mu} p^\mu_\omega
\varepsilon_\nu p^\nu_N + {\bf r}^2\bar u_N u_{*\mu}\varepsilon^\mu
\nonumber\\
& & +4{(m_*+m_N)^2-m^2_\omega\over m_*}\bar u_N \gamma_\nu u_{*\mu}p^\mu_N
\varepsilon^\nu ,
\end{eqnarray}
which give the relation between $g_i$ and $f_i$  vertex form factors:
\begin{eqnarray}
f_1 &=& g_1 + {1\over 3}{\bf r}^2g_2 + {\bf r}^2 g_3 ,\\
f_2 &=& 4{(m_*+m_N)^2-m^2_\omega\over m_*} g_3 \\
f_3 &=& 2(-1+{m^2_N-m^2_\omega\over m^2_*}) g_2
+ 2(-3+{m^2_N-m^2_\omega\over m^2_*}) g_3.
\end{eqnarray}
The $g_i$ and $f_i$ are related by some smooth $m_*$ dependence factors.
For an $N^*$ with very broad width, this may cause some model dependence on
the determination of their mass and width.

\subsection{$\psi\to N^*\bar N$}

Here we give an example of a vector meson decaying into $N^*\bar N$ final state.
\begin{eqnarray}
    &(S_A, S_{BC}, L_{BC})&: \quad -{\bf S_{A}}+{\bf S_{BC}}+{\bf L_{BC}}=0
\nonumber\\
\psi\to N^*({1\over 2}^+)\bar N &(1,1,0)&: \quad
\Psi^{(1)}_\mu\varepsilon^{\mu}, \\
&(1,1,2)&: \quad \Psi^{(1)}_\mu\varepsilon_\nu\tilde t^{(2)\mu\nu} ,\\
\psi\to N^*({1\over 2}^-)\bar N &(1,0,1)&: \quad \psi^{(0)}\varepsilon_\mu\tilde
t^{(1)\mu} ,\\
&(1,1,1)&: \quad 
i\Psi^{(1)}_\mu\epsilon^{\mu\nu\lambda\sigma}\varepsilon_\nu
\tilde t^{(1)}_\lambda\hat p_{(\psi)\sigma}, \\
\psi\to N^*({3\over 2}^+)\bar N &(1,1,0)&: \quad \psi^{(1)}_\mu\varepsilon^{\mu},
\\
&(1,1,2)&: \quad \psi^{(1)}_\mu\varepsilon_\nu\tilde t^{(2)\mu\nu} ,\\
&(1,2,2)&: \quad
i\Psi^{(2)}_{\mu\alpha}\epsilon^{\mu\nu\lambda\sigma}\varepsilon_\nu
\tilde t^{(2)\alpha}_\lambda\hat p_{(\psi)\sigma}, \\
\psi\to N^*({3\over 2}^-)\bar N &(1,1,1)&: \quad
i\psi^{(1)}_\mu\epsilon^{\mu\nu\lambda\sigma}\varepsilon_\nu
\tilde t^{(1)}_\lambda\hat p_{(\psi)\sigma}, \\
&(1,2,1)&: \quad \Psi^{(2)}_{\mu\nu}\varepsilon^{\mu}\tilde t^{(1)\nu} ,\\
&(1,2,3)&: \quad \Psi^{(2)}_{\mu\nu}\varepsilon_\lambda \tilde
t^{(3)\mu\nu\lambda} ,\\
\psi\to N^*({5\over 2}^+)\bar N &(1,2,2)&: \quad
i\psi^{(2)}_{\mu\alpha}\epsilon^{\mu\nu\lambda\sigma}\varepsilon_\nu
\tilde t^{(2)\alpha}_\lambda\hat p_{(\psi)\sigma}, \\
&(1,3,2)&: \quad \Psi^{(3)}_{\mu\nu\lambda}\varepsilon^{\mu}\tilde
t^{(2)\nu\lambda} ,\\
&(1,3,4)&: \quad \Psi^{(3)}_{\mu\nu\lambda}\varepsilon_{\sigma}\tilde
t^{(4)\mu\nu\lambda\sigma} ,\\
\psi\to N^*({5\over 2}^-)\bar N &(1,2,1)&: \quad
\psi^{(2)}_{\mu\nu}\varepsilon^{\mu}\tilde t^{(1)\nu} ,\\
&(1,2,3)&: \quad \psi^{(2)}_{\mu\nu}\varepsilon_\lambda \tilde
t^{(3)\mu\nu\lambda} ,\\
&(1,3,3)&: \quad
i\Psi^{(3)}_{\mu\alpha\beta}\epsilon^{\mu\nu\lambda\sigma}\varepsilon_\nu
\tilde t^{(3)\alpha\beta}_\lambda\hat p_{(\psi)\sigma}, \\
\psi\to N^*({7\over 2}^+)\bar N &(1,3,2)&: \quad
\psi^{(3)}_{\mu\nu\lambda}\varepsilon^{\mu}\tilde t^{(2)\nu\lambda} ,\\
&(1,3,4)&: \quad \psi^{(3)}_{\mu\nu\lambda}\varepsilon_{\sigma}\tilde
t^{(4)\mu\nu\lambda\sigma} ,\\
&(1,4,4)&: \quad
i\Psi^{(4)}_{\mu\alpha\beta\gamma}\epsilon^{\mu\nu\lambda\sigma}\varepsilon_\nu
\tilde t^{(4)\alpha\beta\gamma}_\lambda\hat p_{(\psi)\sigma}, \\
\psi\to N^*({7\over 2}^-)\bar N &(1,3,3)&: \quad
i\psi^{(3)}_{\mu\alpha\beta}\epsilon^{\mu\nu\lambda\sigma}\varepsilon_\nu
\tilde t^{(3)\alpha\beta}_\lambda\hat p_{(\psi)\sigma}, \\
&(1,4,3)&: \quad \Psi^{(4)}_{\mu\nu\lambda\sigma}\varepsilon^{\mu}\tilde
t^{(3)\nu\lambda\sigma} ,\\
&(1,4,5)&: \quad \Psi^{(4)}_{\mu\nu\lambda\sigma}\varepsilon_\delta\tilde
t^{(5)\mu\nu\lambda\sigma\delta}.
\end{eqnarray}
Corresponding couplings in the effective Lagrangian approach are give in
Ref.\cite{Liang}.
In the multipole approach, the amplitude for $\psi\to N^*\bar N$ generally takes the
form 
\begin{equation}
A=\sum_{L,m_L,S,m_S}(L,m_L;S,m_S|1,m_\psi)(S_B,m_B;S_C,m_C|S,m_S) 
Y_{Lm_L}({\bf\hat p_N})G_{LS} |{\bf p_N}|^L f_L(|{\bf p_N}|)
\end{equation}
where $G_{LS}$ is the coupling constant for the final state with orbital angular
momentum L and total spin S, ${\bf p_N}$ is the momentum of $\bar N$ in 
the rest frame of $\psi$ and 
$f_L(|{\bf p_N}|)$ is the vertex form factor. 
Taking $\psi\to N^*({1\over 2}^+)\bar N$ as an example, the amplitude is as the
following
\begin{eqnarray}
\label{LS1}
A &=& ({1\over 2},m_B;{1\over 2}m_C|1,m_\psi) Y_{00}({\bf\hat p_N})G_{01} 
f_0(|{\bf p_N}|) \nonumber\\
&+& (2,m_L;1,m_S|1,m_\psi) ({1\over 2},m_B;{1\over
2}m_C|1,m_S) Y_{2m_L}({\bf\hat p_N})G_{21}|{\bf p_N}|^2 f_2(|{\bf p_N}|)
\end{eqnarray}
with $m_S=m_B+m_C$ and $m_L=m_\psi-m_S$.
With some simple algebra, the corresponding amplitude in the
covariant L-S scheme can be reduced to the similar form:
\begin{eqnarray}
\label{LS2}
A &=& g_0\Psi^{(1)}_\mu\varepsilon^{\mu}f_0(|{\bf p_N}|)+
g_2\Psi^{(1)}_\mu\varepsilon_\nu\tilde t^{(2)\mu\nu}f_2(|{\bf p_N}|) \nonumber\\
&=& ({1\over 2},m_B;{1\over 2}m_C|1,m_\psi) Y_{00}({\bf\hat
p_N}) g_0\sqrt{8\pi}C_\Psi f_0(|{\bf p_N}|) \nonumber\\
& & +(2,m_L;1,m_S|1,m_\psi) ({1\over 2},m_B;{1\over 2}m_C|1,m_S) Y_{2m_L}({\bf\hat
p_N})g_2{8\over 3}\sqrt{4\pi}C_\Psi |{\bf p_N}|^2 f_2(|{\bf p_N}|). \nonumber\\
\end{eqnarray}
Comparing Eq.(\ref{LS1}) and Eq.(\ref{LS2}), we have
\begin{eqnarray}
G_{01} &=& g_0 \sqrt{8\pi}C_\Psi ,\\
G_{21} &=& g_2 {8\over 3}\sqrt{4\pi}C_\Psi. 
\end{eqnarray}
In non-relativistic limit, $C_\Psi=1$, and the covariant L-S scheme gives
$G_{01}$ and $G_{21}$ as constants; but generally speaking,  the covariant L-S
scheme results in $G_{01}$ and $G_{21}$ smoothly dependent on $|\bf p_N|$. 

As a concrete example, here we study the angular distribution and
the relative ratio of D-wave and S-wave in the final states of
$e^+e^-\to J/\psi \to p \bar p$. For this process of positron-electron
collision, $J/\psi$ spin projection is limited to be $\pm 1$ along
the beam direction. The differential decay rate of the $J/\psi$ is related 
to the amplitude A as
\begin{equation}
{d\Gamma\over d\Omega} = {1\over 32\pi^2} \overline{|A|^2} 
{|{\bf p_N}|\over M^2_\psi} .
\end{equation}
With $A$ given by Eq.(97), we have
\begin{equation}
\overline{|A|^2}={m^2_\psi\over m^2_p}({1\over 2}C^2_S
+{20\over 9}C^2_D-{2\over 3}C_SC_Dcos\beta)
(1+\alpha cos^2\theta)
\end{equation} 
where
\begin{equation}
\alpha=\frac{2C_SC_Dcos\beta-{4\over 3}C^2_D}
{{1\over 2}C^2_S+{20\over 9}C^2_D-{2\over 3}C_SC_Dcos\beta}
\end{equation}
with $C_S\equiv |g_0|f_0(|{\bf p_N}|)$, 
$C_D\equiv |g_2|{\bf p_N^2}f_2(|{\bf p_N}|)$ 
and $\beta$ the relative phase between $C_S$ and $C_D$.
If $C_D=0$, then $\alpha=0$ as expected for a pure S-wave decay;
if $C_S=0$, then $\alpha=-{3\over 5}$ for the pure D-wave decay. 

The relative ratio $R_{D/S}$ of D-wave and S-wave decay rates is
\begin{equation}
R_{D/S}\equiv {\Gamma_D\over\Gamma_S}={32C^2_D\over C^2_S}.
\end{equation}

The experimental value of $\alpha$ for the $e^+e^-\to J/\psi \to p \bar p$
process is about 0.62\cite{DM2}. This gives the  ratio $R_{D/S}$ to be
in the range of $0.09\sim 1.9$. The large uncertainty is due to the 
unknown relative phase $\beta$ between S-wave and D-wave amplitudes. For a 
full determination of the  ratio $R_{D/S}$, the polarization information 
of final state particles is needed.

\section{Discussion}

Comparing with the simple effective Lagrangian approach, each coupling in
the covariant L-S scheme corresponds to a single L final states while
a coupling in the  simple effective Lagrangian approach may be a mixture
of two L final states. The number of independent couplings is same
in the two approaches as it should be. In the simple effective Lagrangian approach,
the independent couplings are not necessary to be orthogonal to each other;
while in the covariant L-S scheme, they are orthogonal and make the partial wave
analysis easier.
The construction of the full
amplitude in the covariant L-S scheme for a multi-step process, {\sl e.g.},
$J/\psi\to N^*\bar N\to \omega N\bar N$, is similar to the simple effective
Lagrangian approach\cite{Liang}. The coupling constants for each couplings
are fitted to the data in the procedure of partial wave analysis\cite{BES1}.

For the partial wave analysis, we only demand very basic
requirements, {\sl i.e.}, Lorentz, CPT, C and P invariance, 
for the amplitude and we make formalism more general. Various
theories or models or assumptions can bring more constraints to
the relations of various couplings, hence reduce the number of
independent couplings. For example, a chiral quark model
calculation\cite{Riska} results in a single coupling form for the
$N^*(1675)({5\over 2}^-) N\omega$ coupling, which corresponds to
our (1,2,2) coupling of Eq.(55), while other quark model\cite{Capstick2} gives
different prediction. This can be checked in the future
by partial wave analysis of processes involving $N^*(1675)({5\over
2}^-)\to N\omega$. Some people\cite{Lee} assume $N^*({3\over
2}^\pm)N\omega$ couplings to have the same structure as
$N^*({3\over 2}^\pm)N\gamma$ hence only two independent couplings.
In our general scheme, we have three independent couplings for
$N^*({3\over 2}^\pm)N\omega$ couplings; gauge invariance
requirement for the $N^*({3\over 2}^\pm)N\gamma$ couplings reduces
the number of independent couplings to two for the $N^*({3\over
2}^\pm)N\gamma$ couplings.

In this paper we have given explicit formulae for $N^*\to N\pi$,
$N^*\to N\omega$ and $\psi\to N^*\bar N$ as examples since the
relevant processes are understudy by experimental groups.
 For any baryon resonance
decaying to a ${1\over 2}^+$ baryon plus a pseudoscalar meson
through strong interaction, {\sl e.g.}, $N^*\to\Lambda K$,
$N^*\to\Sigma K$, $\Lambda^*\to NK$, $\Lambda^*\to\Sigma\pi$, {\sl
etc}., the coupling has the same form as for $N^*\to N\pi$, the
only difference is the coupling constants. For any baryon
resonance strong decaying to a ${1\over 2}^+$ baryon plus a vector
meson, the coupling has the same form as for $N^*\to N\omega$. For
any vector meson strong decaying to a baryon resonance plus an
anti-(${1\over 2}^+$)baryon, the coupling has the same form as for
$\psi\to N^*\bar N$. Extension to other processes are
straightforward by following the basic rules outlined in this
work.

In our present L-S scheme for $N^*$ decays, we have added the spin of
the incoming nucleon resonance and the final nucleon. This is different 
with the usual L-S scheme where it is always the spin of the final state 
particles which are added to make the total spin S. The two schemes are
simply related by recoupling various angular momenta involved.
With recoupling technique in Ref.\cite{Lawson}, we have the relation 
between the two schemes as the following:
\begin{eqnarray}
\left[ [{\bf S}_A\times {\bf S}_B]_{S_{AB}}\times {\bf S}_C\right]_{LM}
&=&\sum_{S_{BC}}\sqrt{(2S_{AB}+1)(2S_{BC}+1)}W(S_AS_BLS_C;S_{AB}S_{BC})
\nonumber\\
&& \quad\quad \left[ {\bf S}_A\times [{\bf S}_B\times {\bf 
S}_C]_{S_{BC}}\right]_{LM}
\end{eqnarray} 
where $W(S_AS_BLS_C;S_{AB}S_{BC})$ is the usual Racah 
coefficients\cite{Lawson}.
From this relation, after we get the coupling constants in our scheme,
$g(S_{AB},L)$, we can easily get the corresponding coupling constants
in the usual L-S scheme, $G(S_{BC},L)$, as
\begin{equation}
G(S_{BC},L)=\sum_{S_{AB}}g(S_{AB},L)\sqrt{(2S_{AB}+1)(2S_{BC}+1)}
W(S_AS_BLS_C;S_{AB}S_{BC}).
\end{equation}

Since the covariant L-S scheme combines merits of  
two conventional schemes, {\sl i.e.}, covariant effective Lagrangian approach    
and multipole analysis with amplitudes expanded according to  angular momentum L,
we recommend it to be used in future partial wave analysis.

\bigskip
{\bf Acknowledgements:} We thank D.V.Bugg, J.G. K\"orner, T.S.H.Lee, 
P.N.Shen, J.X.Wang and J.J.Zhu
for useful discussions. This work is partly supported by the National Science
Foundation of China and by the CAS Knowledge Innovation Project KJCX2-SW-N02.

\end{document}